\documentclass[sn-mathphys,Numbered,]{sn-jnl}
\usepackage{graphicx}%
\usepackage{multirow}%
\usepackage{amsmath,amssymb,amsfonts}%
\usepackage{amsthm}%
\usepackage{mathrsfs}%
\usepackage[title]{appendix}%
\usepackage{xcolor}%
\usepackage{textcomp}%
\usepackage{manyfoot}%
\usepackage{booktabs}%
\usepackage{algorithm}%
\usepackage{algorithmicx}%
\usepackage{algpseudocode}%
\usepackage{listings}%
\usepackage[numbers]{natbib}

\theoremstyle{thmstyleone}%
\theoremstyle{thmstyletwo}%

\theoremstyle{thmstylethree}%

\newcommand{\be}{\begin{equation}}
\newcommand{\ee}{\end{equation}}

\newcommand{\fedd}{f_{\rm Edd}}

\newcommand{\Ledd}{L_{\rm Edd}}

\newcommand{\mpro}{m_{\rm p}}

\begin{document}

\title[Article]{Formation and Growth of the First Supermassive Black Holes in MOG}

\author*[1,2,3]{\fnm{Mohammad H. } \sur{Zhoolideh Haghighi}}\email{zhoolideh@kntu.ac.ir}

\author[3,4]{\fnm{John} \sur{Moffat}}\email{jmoffat@perimeterinstitute.ca}

\affil*[1]{\orgdiv{Department of Physics}, \orgname{K.N. Toosi University of Technology}, \orgaddress{ \city{Tehran}, \postcode{P.O. Box 15875-4416}, \state{Tehran}, \country{Iran}}}

\affil[2]{\orgdiv{School of Astronomy}, \orgname{Institute for Research in Fundamental Sciences (IPM)}, \orgaddress{\city{Tehran}, \postcode{19395–5746}, \state{Tehran}, \country{Iran}}}

\affil[3]{\orgdiv{Perimeter Institute for Theoretical Physics}, \orgaddress{\city{Waterloo}, \postcode{N2L 2Y5}, \state{Ontario}, \country{Canada}}}

\affil[4]{\orgdiv{Department of Physics and Astronomy}, \orgname{University of Waterloo}, \orgaddress{\city{Waterloo}, \postcode{N2L 3G1}, \state{Ontario}, \country{Canada}}}


\abstract{The emergence of supermassive black holes (SMBHs) in the early universe remains a topic of profound interest and debate. In this paper, we investigate the formation and growth of the first SMBHs within the framework of Modified Gravity (MOG), where gravity exhibits increased strength. We explore how MOG, as an alternative to the standard model, may offer novel insights into the emergence of SMBHs and potentially reconcile the discrepancies observed in the accretion and growth processes. We examine the dynamics of gas and matter in this modified gravitational framework, shedding light on the unique interplay between gravity and the formation of SMBHs.}


\maketitle


\section{Introduction}

The existence of supermassive black holes at the centers of galaxies is a well-established astrophysical phenomenon. However, the formation and growth of these enigmatic objects, particularly in the context of the early universe, remain subjects of intense scrutiny and debate. Traditional models, relying on the framework of General Relativity and the presence of dark matter, have provided some insights into the growth mechanisms of SMBHs over cosmic time. Nevertheless, discrepancies persist in our understanding of the accretion of black hole seeds and their rapid growth observed within approximately the first 800 million years after the Big Bang. Specifically, the detection of SMBHs with masses exceeding $10^9$ $M_{\odot}$ at redshifts z $\ge 7$ \cite{2011Natur.474..616M, 2006AJ....131.1203F} has presented a significant challenge to the theories concerning the origin and enlargement of black holes. This observation, documented in studies by \cite{2010AJ....139..906W, 2011Natur.474..616M, 2018Natur.553..473B, 2001AJ....122.2833F,2020ARA&A..58...27I, 2019ffbh.book.....L}, and current observations of the JWST telescope, which have detected a massive BH with $M_{BH} \sim 10^7 - 10^8 M_{\odot}$ at $z \sim 10.3$ \cite{2023arXiv230515458B}, has raised profound questions in the field of astrophysics.

Traditionally, black holes are anticipated to originate as the ultimate outcome of massive star evolution \cite{2013IAUS..292..184W, 2001AJ....122.2833F, 2019MNRAS.484.2575T, 2002RvMP...74.1015W}.
Researchers have explored various avenues to gain insights into the appearance of SMBHs when the Universe was just $\sim 800~ Myr$. These scenarios have been mainly categorized into two groups of light seed scenarios and heavy seed scenarios \cite{2019MNRAS.486.3892R}.
Light seed scenarios encompass mechanisms where the initial mass of the black hole is relatively low, in which the initial mass is around 100 $M_{\odot}$, but rapid growth happens. In the standard light seeding models, the SMBH starts growing from a Pop III stellar remnant \cite{2001ApJ...551L..27M}, and it has been shown that this growth favors MHD accretion disks over standard thin disks \cite{2005ApJ...620...59S}.
Alternatively, there are scenarios where a heavy seed, with an initial mass greater than approximately 10,000 $M_{\odot}$, but a smaller growth rate could come into existence. Within the cores of rapidly accreting atomic cooling haloes devoid of metals, it is expected that a supermassive star (SMS) may form. The final mass of the SMS is projected to greatly exceed 10,000 $M_{\odot}$. This collapse into a direct collapse black hole yields a black hole seed with a substantial initial mass. In addition to this, Bondi accretion, together with super-Eddington gas accretion for black holes can be an attractive way to solve the rapid growth problem of SMBHs \cite{2020arXiv201113440M}.

A probable solution for the mentioned growth problem can be via enhancing the gravitational force. Some authors do it via using dark matter halos \cite{2023arXiv231006898S}, but some prefer to solve the problem through the Modified Gravity (MOG) framework \cite{Moffat2006}.
MOG is a theoretical framework proposing modifications to the laws of gravity in which gravity is posited to exhibit increased strength, departing from the predictions of General Relativity. MOG introduces a scalar field gravitational coupling strength $G$ and a gravitational spin 1 vector field $\phi_\mu$. The $G$ is written as $G=G_N(1+\alpha)$ where $G_N$ is Newton's constant, and the gravitational source charge for the vector field is $Q_g=\sqrt{\alpha G_N}M$, where $M$ is the mass of a body. MOG has been successful in fitting observations in different scales including dwarf galaxies \cite{2017MNRAS.468.4048Z}, spiral galaxies \cite{2013MNRAS.436.1439M}, clusters of galaxies \cite{2016arXiv161105382M, 2014MNRAS.441.3724M}, and cosmology \cite{2021MNRAS.507.3387D, 2021Univ....7..358M}, without any need for including dark matter. We will discuss how MOG can provide an alternative solution for SMBH formation growth, raising the possibility of novel insights that may bridge the gaps in our understanding without the use of dark matter.

This paper aims to investigate the formation and growth of the first SMBHs within the MOG paradigm to see if MOG can offer a viable explanation for the existence of supermassive black holes at redshifts $z\ge7$. We begin in section \ref{section:MOG} of this manuscript with a brief explanation of the MOG theory. In section \ref{section:MOG BH} we introduce MOG black holes. In section \ref{section:smbh growth} we explain supermassive black hole growth and evolution; adapt the calculation to the MOG theory, and show the results. We conclude by presenting our discussion and conclusion in section \ref{section:conclusions}.

\section{Modified Gravity (MOG)} \label{section:MOG}

We use the version of MOG action and field equations formulation proposed in \cite{Moffat2}, in which $\chi= 1/G$ where $\chi$ is a scalar field and $G$ is the coupling strength of gravity. The MOG action is
given by (we use the metric signature $(+,-,-,-)$ and units with $c=1$): \be S=S_G+S_\phi+S_M, \ee where \be S_G=\frac{1}{16\pi}\int
d^4x\sqrt{-g}\biggl(\chi R+\frac{\omega_M}{\chi}\nabla^\mu\chi\nabla_\mu\chi +2\Lambda\biggr), \ee and \be \label{Baction} S_\phi=\int
d^4x\sqrt{-g}\biggl(-\frac{1}{4}B^{\mu\nu}B_{\mu\nu}+\frac{1}{2}\mu^2\phi^\mu\phi_\mu\biggr). \ee $S_M$ is the matter action and $J^\mu$ is the
current matter source of the vector field $\phi_\mu$. Moreover, $\nabla_\mu$ denotes the covariant derivative with respect to the metric
$g_{\mu\nu}$, $B_{\mu\nu}=\partial_\mu\phi_\nu-\partial_\nu\phi_\mu$ and $\omega_M$ is a constant. The Ricci tensor is \be
R_{\mu\nu}=\partial_\lambda\Gamma^\lambda_{\mu\nu}-\partial_\nu\Gamma^\lambda_{\mu\lambda}
+\Gamma^\lambda_{\mu\nu}\Gamma^\sigma_{\lambda\sigma}-\Gamma^\sigma_{\mu\lambda}\Gamma^\lambda_{\nu\sigma}. \ee We expand $G$ by
$G=G_N(1+\alpha)$, $\Lambda$ is the cosmological constant and $\mu$ is the effective running mass of the spin 1 graviton vector field.

Variation of the matter action $S_M$ yields \be T^M_{\mu\nu}=-\frac{2}{\sqrt{-g}}\frac{\delta S_M}{\delta g^{\mu\nu}},\quad
J^\mu=-\frac{1}{\sqrt{-g}}\frac{\delta S_M}{\delta\phi_\mu}. \ee Varying the action with respect to $g_{\mu\nu}$, $\chi$ and $\phi_\mu$, we
obtain the field equations: \be \label{Gequation} G_{\mu\nu}=-\frac{\omega_M}{\chi^2}\biggl(\nabla_\mu\chi\nabla_\nu\chi
-\frac{1}{2}g_{\mu\nu}\nabla^\alpha\chi\nabla_\alpha\chi\biggr)\\
-\frac{1}{\chi}(\nabla_\mu\chi\nabla_\nu\chi-g_{\mu\nu}\Box\chi)+\frac{8\pi}{\chi}T_{\mu\nu}, \ee \be \label{Bequation} \nabla_\nu
B^{\mu\nu}+\mu^2\phi^\mu=J^\mu, \ee \be \label{Boxchi} \Box\chi=\frac{8\pi}{(2\omega_M+3)}T, \ee where $G_{\mu\nu}=R_{\mu\nu}-\frac{1}{2}R$,
$\Box=\nabla^\mu\nabla_\mu$, $J^\mu=\kappa\rho_M u^\mu$, $\kappa=\sqrt{G_N\alpha}$, $\rho_M$ is the density of matter and $u^\mu=dx^\mu/ds$. The
energy-momentum tensor is \be T_{\mu\nu}=T^M_{\mu\nu}+T^\phi_{\mu\nu}+g_{\mu\nu}\frac{\chi\Lambda}{8\pi}, \ee where \be
T^\phi_{\mu\nu}=-\biggl({B_\mu}^\alpha
B_{\alpha\nu}-\frac{1}{4}g_{\mu\nu}B^{\alpha\beta}B_{\alpha\beta}+\mu^2\phi_\mu\phi_\nu-\frac{1}{2}g_{\mu\nu}\phi^\alpha\phi_\alpha\biggr), \ee
and $T=g^{\mu\nu}T_{\mu\nu}$. Eq.(\ref{Boxchi}) follows from the equation \be 2\chi\Box\chi-\nabla_\mu\chi\nabla^\mu\chi=\frac{R}{\omega_M}, \ee
by substituting for $R$ from the contracted form of (\ref{Gequation}).

\section{MOG Black Hole} \label{section:MOG BH}

For the matter-free $\phi_\mu$ field-vacuum case with $\Lambda=0$, $T^M_{\mu\nu}=0$ and $J^\mu=0$, the field equations are given by
\be
G_{\mu\nu}=-\frac{\omega_M}{\chi^2}\biggl(\nabla_\mu\chi\nabla_\nu\chi -\frac{1}{2}g_{\mu\nu}\nabla^\alpha\chi\nabla_\alpha\chi\biggr)\\
-\frac{1}{\chi}(\nabla_\mu\chi\nabla_\nu\chi-g_{\mu\nu}\Box\chi)+\frac{8\pi}{\chi}T^\phi_{\mu\nu},
\ee
\be
\label{Boxphivac}
\Box\chi=\frac{8\pi}{(2\omega_M+3)}T^\phi,
\ee
\be \nabla_\nu B^{\mu\nu}+\mu^2\phi^\mu=0, \ee where \be T^\phi\equiv
g^{\mu\nu}T^\phi_{\mu\nu}=\mu^2\phi^\mu\phi_\mu.
\ee
As shown in~\cite{Moffat3} for the black hole solution, the vector field effective mass $\mu$ can be set to zero.

The conformally invariant gravitational $\phi$-field energy-momentum tensor is
\be
\label{Bfieldenergy} T^\phi_{\mu\nu}=-\biggl({B_\mu}^\alpha
B_{\alpha\nu}-\frac{1}{4}g_{\mu\nu}B^{\alpha\beta}B_{\alpha\beta}\biggr),
\ee
and we have $T^\phi=g^{\mu\nu}T^\phi_{\mu\nu}=0$ giving the
equation:
\be
\label{Boxchi2} \Box\chi=0.
\ee
We now multiply (\ref{Boxchi2}) by $\chi$ and integrate over the volume ${\cal V}$. Integrating by parts, we derive the volume integral~\cite{Hawking}:

\be
\label{gradientchi} \int_{\cal V}d^4x\sqrt{-g}\nabla^\mu\chi\nabla_\mu\chi=\int_{\partial{\cal
V}}d^3x\sqrt{|h|}\nabla_\mu\chi n^\mu=0.
\ee
Because the volume integral (\ref{gradientchi}) must be zero and the gradient of $\chi$ is either
spacelike or zero, then the gradient of $\chi$ must be zero everywhere and $\chi$ must be constant. In the original papers on MOG black
holes~\cite{Moffat2,Moffat3}, we assumed that the gradient of $G$ is zero, so that $G=G_N(1+\alpha)=1/\chi$ is constant. We can now demonstrate
that for MOG black holes, it is justified to have $G$ constant.

The metric for a static spherically symmetric black hole is given by~\cite{Moffat2,Moffat3}:
\be
\label{MOGmetric}
ds^2=\biggl(1-\frac{2G_N(1+\alpha)M}{r}+\frac{\alpha(1+\alpha)G_N^2M^2}{r^2}\biggr)dt^2-\biggl(1-\frac{2G_N(1+\alpha)M}{r}+
\frac{\alpha(1+\alpha)G_N^2M^2}{r^2}\biggr)^{-1}dr^2-r^2d\Omega^2,
\ee
where $d\Omega^2=d\theta^2+\sin^2\theta d\phi^2$. The metric reduces to
the Schwarzschild solution when $\alpha=0$. The axisymmetric rotating black hole has the metric solution including the spin angular momentum
$J=G_NM^2 a/c$ where $a$ is the dimensionless spin parameter:
\be
ds^2=\frac{\Delta}{\rho^2}(dt-a\sin^2\theta
d\phi)^2-\frac{\sin^2\theta}{\rho^2}[(r^2+a^2)d\phi-adt]^2-\frac{\rho^2}{\Delta}dr^2-\rho^2d\theta^2,
\ee
where
\be
\Delta=r^2-2G_N(1+\alpha)Mr+a^2+\alpha(1+\alpha)G_N^2M^2,\quad \rho^2=r^2+a^2\cos^2\theta.
\ee
Consider the timelike Killing vector $\xi^\mu$ in a stationary asymptotically flat solution of an uncollapsed body. The gravitational mass of
the body measured at infinity is given by the $1/r$ and the $1/r^2$ contributions to the MOG metric in $\xi^\mu\xi^\nu g_{\mu\nu}$. We have \be
M=\frac{1}{4}\pi\int d\Sigma_{\nu\beta}\nabla_\mu xg^{\mu\nu}\xi^\beta, \ee where $\Sigma_{\nu\beta}$ is the surface element of a spacelike
2-surface at infinity and $x^2=\xi^\mu\xi^\nu g_{\mu\nu}$. In the Einstein frame, $M$ denotes the gravitational mass calculated in the Einstein
frame. Because the source mass $M$ in the gravitational charge of the vector field, $Q_g=\sqrt{\alpha G_N}M$, is positive, the vector (spin 1
graviton) field does not produce a dipole moment. A consequence is that for an isolated black hole, there will not be any dipole vector
gravitational source, and because $\chi=1/G$ is constant there will not be any scalar monopole gravitational source. It can be demonstrated that
for merging black holes $M_E$ will decrease by the amount of tensor (quadrupole) field energy radiated at infinity.

\section{Supermassive Black Hole Growth and Evolution}\label{section:smbh growth}

The SMBH in active galactic nuclei and quasars span a mass range, extending up to nearly $10^{11}M_{\odot}$~\cite{McConnell2011}. The heaviest black hole is associated with the quasar TON 618 with a mass $\sim 7\times 10^{10}M_{\odot}$~\cite{Shemmer2004}, while the second heaviest in the galaxy IC 1101 has a mass $\sim 4\times 10^{10}M_{\odot}$~\cite{Dullo2017}. This raises the question of whether there is enough time for these objects to be formed. For an SMBH to grow to a black hole with a mass $10^9-10^{10}M_{\odot}$ in the short time after the Big Bang, and to account for the short lifetime of detected quasars, several scenarios have been proposed.

Some suggest a rapid accretion scenario or continuous accretion. The key to rapid SMBH growth is exceptionally high rates of accretion, where the black hole continuously gathers mass from its surrounding environment ~\cite{2006MNRAS.370..289B}. In the early universe, there were regions with abundant gas and dust that could provide the necessary fuel for accretion. The availability of dense gas and dust in its vicinity, coupled with its powerful gravitational pull, allows it to accumulate mass rapidly ~\cite{Volonteri2010}. It is worth mentioning that the high-density gas required for continuous accretion may not be readily available in all regions of the early universe. Also, feedback mechanisms from the growing SMBH can disrupt accretion~\cite{2005Natur.433..604D}.
Moreover, this idea has been criticized for the difficulty in achieving the extremely high accretion rates required to grow a 100 $M_{\odot}$ seed into a billion solar mass SMBH within the available time frame~\cite{2012Sci...337..544V}. One alternative hypothesis is that the 100 $M_{\odot}$ seed rapidly accretes gas and grows into a supermassive star, sometimes referred to as a supermassive star seed or direct collapse star. These stars can reach thousands of solar masses due to their rapid accretion rates~\cite{Begelman2006}. While the concept of supermassive star formation leading to a black hole is theoretically possible, the formation and stability of such massive stars are subjects of debate. The main challenge is understanding how these stars can avoid fragmentation and continue to accrete mass~\cite{2012Sci...337..544V}. Finally, there is a possibility to resolve the problem via feedback mechanisms. The intense radiation and energy emitted by the growing black hole can influence its surroundings. This can suppress the formation of new stars in the vicinity, preventing their radiation from disrupting the accretion process~\cite{2005Natur.433..604D}. Nevertheless, some studies show that the feedback from active SMBHs, such as quasars, can impact the surrounding environment, potentially hindering the growth of the SMBH itself~\cite{Shankar2009}.

One possible alternative solution to tackle the problem could be through MOG. To investigate how MOG can solve the issue, first let us find the black hole mass as a function of time. For spherical infall of gas on a black hole, we have to account for the Eddington limit on luminosity. For accretion on a mass $M$, if the gravitational attractive force does not exceed the radiation gas pressure, the luminosity $L$ cannot exceed $L_{\rm Edd}$.
So, the initial model for black hole accretion is described by the Eddington rate \citep{Frank2002}. This rate can be derived by equating the gravitational force acting inward on each proton to the radiation pressure exerted outward on its neighboring electron, which is positioned at a distance $r$ from the central point.
\be
\Ledd = \frac{4 \pi c G \mpro M}{\sigma_T}.
\label{Ledd}
\ee
Here $\sigma_T=\frac{8\pi/3}{(e^2/m_e c^2)^2}$ is the Thomson scattering cross section and $m_p$ is the proton mass. When the black hole is undergoing accretion at a rate denoted as $\dot{M}$, a portion of the gravitational potential energy can be emitted as radiation. If we represent this as a fraction of the rest-mass energy, the resulting radiated luminosity can be expressed as follows:
\be
L_{acc} = \epsilon \dot{M}_{acc} c^2,
\label{lumin}
\ee
where $\epsilon$ is radiative efficiency, whose value has been observationally constrained to be about 0.1~\cite{2017SCPMA..60j9511Z}, and $\dot{M}$ is mass accretion rate. $\epsilon$ plays a very significant role in the mass accretion of SMBHs and small changes in its value can result in significant differences in mass amplification by accretion at the Eddington
limit \cite{2005ApJ...620...59S}.
Using $\dot{M} = (1-\epsilon)\dot{M}_{acc}$, Eddington fraction, $f_{Edd} = \frac{L_{acc}}{L_{Edd}}$, and Eq.(\ref{lumin}) and (\ref{Ledd}), one can
find the black hole mass as a function of time.
\be
\label{mass-rate}
\frac{dM}{dt} =  \gamma_0\, f_{Edd} \, M \Rightarrow M(t) = M_0 \exp(\gamma_0 \, f_{Edd} \, t),
\ee
where $\gamma_0 = (1-\epsilon)/(\epsilon t_E)$ and the Salpeter/Eddington time is, $t_E = 2e^4/(3G_N m_p m_e^2c^3) = 450$ Myr.
When it comes to MOG, Eq.(\ref{mass-rate}) should be modified as the gravitational constant $G_N$ needs to be replaced with $G = (1+\alpha)G_N$. As a result, we have:
\be
\label{mog-mass-rate}
\frac{dM}{dt} = \gamma_0 \, f_{Edd} \, M \Rightarrow M(t) = M_0 \exp((1+\alpha)\, f_{Edd}\, \gamma_0 \, t).
\ee
MOG increases the gravitational force by a $(1+\alpha)$ factor, so particles experience stronger gravity.
We note that the angular momentum of rotation or spin for a steady-state black hole accretion, determined by the spin parameter $a=Jc/M^2G_N$ is related to the radiative efficiency $\epsilon$~\cite{2021ARA&A..59..117R}. In fact, $\epsilon$  is a function of spin and it increases as spin increases. When we account for the MOG enhancement of the gravitational constant, $G=G_N(1+\alpha)$, we decrease the range of the spin parameter, $a_{MOG}=Jc/M^2G_N(1+\alpha) < a$ ~ for ~ $\alpha > 0$. This will increase the infall of matter into the black hole and increase the rate of accretion of matter.

\begin{figure}
\centering
\includegraphics[scale=1.5]{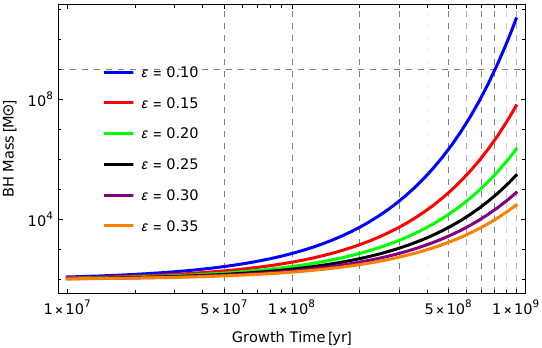}
\caption{Black hole mass as a function of time in standard GR for $f_{Edd} = 1.0$ and different values of $\epsilon$. The horizontal dashed line indicates the $10^{9}M_{\odot}$.}
\label{mass_time}
\end{figure}

\begin{figure}
\centering
\includegraphics[scale=1.5]{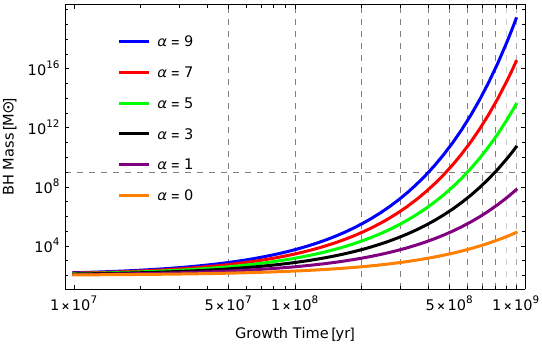}
\caption{Black hole mass as a function of time in MOG for $\epsilon = 0.3$, $f_{Edd} = 1.0$, and different values of $\alpha$. The horizontal dashed line indicates the $10^{9}M_{\odot}$.}
\label{mass_time-mog}
\end{figure}

Therefore, an 800 Myr time period, which could be insufficient for standard black holes to turn into a SMBH, is more than enough time for MOG BHs to accumulate $10^{9}M_{\odot}$ mass. In Fig.(\ref{mass_time}), we have plotted the BH mass as a function of time for  $f_{Edd} = 1$, and $M_0 = 100 M_{\odot}$ for standard BHs. As can be seen the smaller the value of  $\epsilon$, the faster the BH accretes mass. However, none of these epsilons is able to accrete $10^{9}M_{\odot}$ in 800 Myr years. It is improbable for black holes to accrete continuously with $f_{Edd}=1$ throughout all stages of their lifetime. However, this is not an issue for MOG as there is the $\alpha$ parameter, which can strengthen the gravity and accrete the necessary mass within the required time frame. This can be seen from Fig.(\ref{mass_time-mog}), which is plotted for $\epsilon = 0.30$,  $f_{Edd} = 1$, $M_0 = 100 M_{\odot}$, and various values of $\alpha$. For values bigger than $\alpha=3$, SMBHs with mass $10^{9}M_{\odot}$ can be easily formed within 800 Myrs.

\begin{figure}[h]%
\centering
\includegraphics[scale=1.2]{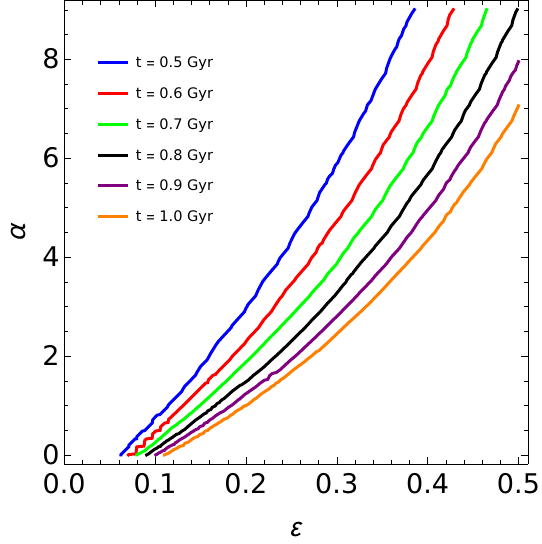}
\caption{$\alpha-\epsilon$  contours for various values of time necessary to gain $10^{9}M_{\odot}$ SMBHs. Here we have fixed $\fedd = 1.0$.}
\label{alpha-eps}
\end{figure}

\begin{figure}
\centering
\includegraphics[scale=1.2]{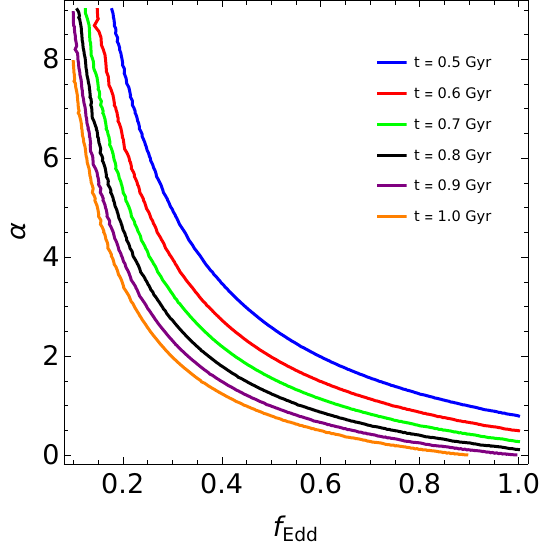}
\caption{$\alpha-\fedd$  contours for various values of time necessary to gain $10^{9}M_{\odot}$ SMBHs. Here we have fixed $\epsilon = 0.1$.}
\label{alpha-fedd}
\end{figure}

Moreover, it can be seen from Eq.(\ref{mog-mass-rate}), that $\alpha$, $\epsilon$, and $f_{Edd}$ play roles in the determination of mass accretion. We have illustrated this in contour plots of Fig.(\ref{alpha-eps}), in which we have used $f_{Edd} = 1$ in Fig.(\ref{alpha-eps}), and $\epsilon = 0.1$ in Fig.(\ref{alpha-fedd}). These contours show the required values of  $\alpha$, $\epsilon$, and $f_{Edd}$ to get SMBHs of mass $10^{9}M_{\odot}$ at the different given times. These plots show that for having a SMBH in times between 500 million to one billion years, one has to either choose standard GR with small $\epsilon$ or select a bigger $\epsilon$ in the modified gravity (MOG) framework. It can be inferred from Figs.(\ref{alpha-eps} \& \ref{alpha-fedd}) that even if the future observations determine $\epsilon > 0.1$, or $f_{Edd} < 1$, MOG can provide an explanation for SMBH formation, taking advantage of the $\alpha$ parameter. This can be considered as an observational test of MOG as well, and future observations of radiative efficiency will determine whether MOG is needed to explain the growth of SMBHs in the early universe.

In order to see whether the age of the universe for MOG could be different in various redshifts and also to have a comparison with the results of \cite{2005ApJ...620...59S}, we obtain the time parameter as a function of redshift and calculate the black hole accretion mass amplification $(M_f/M_i)$ in terms of $z$. The scale factor $a(t)$ in MOG for the spatially flat universe (k = 0) is evolved according to the following equation~\cite{2015arXiv151007037M}:

\begin{equation} \label{FRW}
H^2 \equiv \left( \frac{\dot a}{a} \right)^2 =
H_0^2 \left[ \Omega^0_m \left( \frac{a_0}{a} \right)^3 +
     \Omega^0_\Lambda \right],
\end{equation}
where H(t) is the Hubble constant and $ \Omega^0_m=\Omega^0_b+\Omega^0_{DM}=(1+\alpha)~\Omega^0_b $. The subscript and superscript “0” denotes the value of a quantity at z = 0. Using the fact that $\Omega^0_m + \Omega^0_\Lambda = 1$ with $a/a_0 = 1+z$ and integrating Eq. (\ref{FRW}), we have

\begin{equation} \label{z}
t(z) = \frac{2}{3 H_0 (1-\Omega_m^0)^{1/2}}
{\rm sinh}^{-1}\left[\left (\frac {1-\Omega_m^0}{\Omega_m^0} \right)
\frac{1}{(1+z)^3} \right]^{1/2}.
\end{equation}
We use Planck values for the cosmological parameters: $\Omega^0_m = 0.315$ and $H_0 = 67.4$ $km ~s^{-1} Mpc^{-1}$~\cite{2020A&A...641A...6P}.

\begin{figure}
\centering
\includegraphics[scale=0.75]{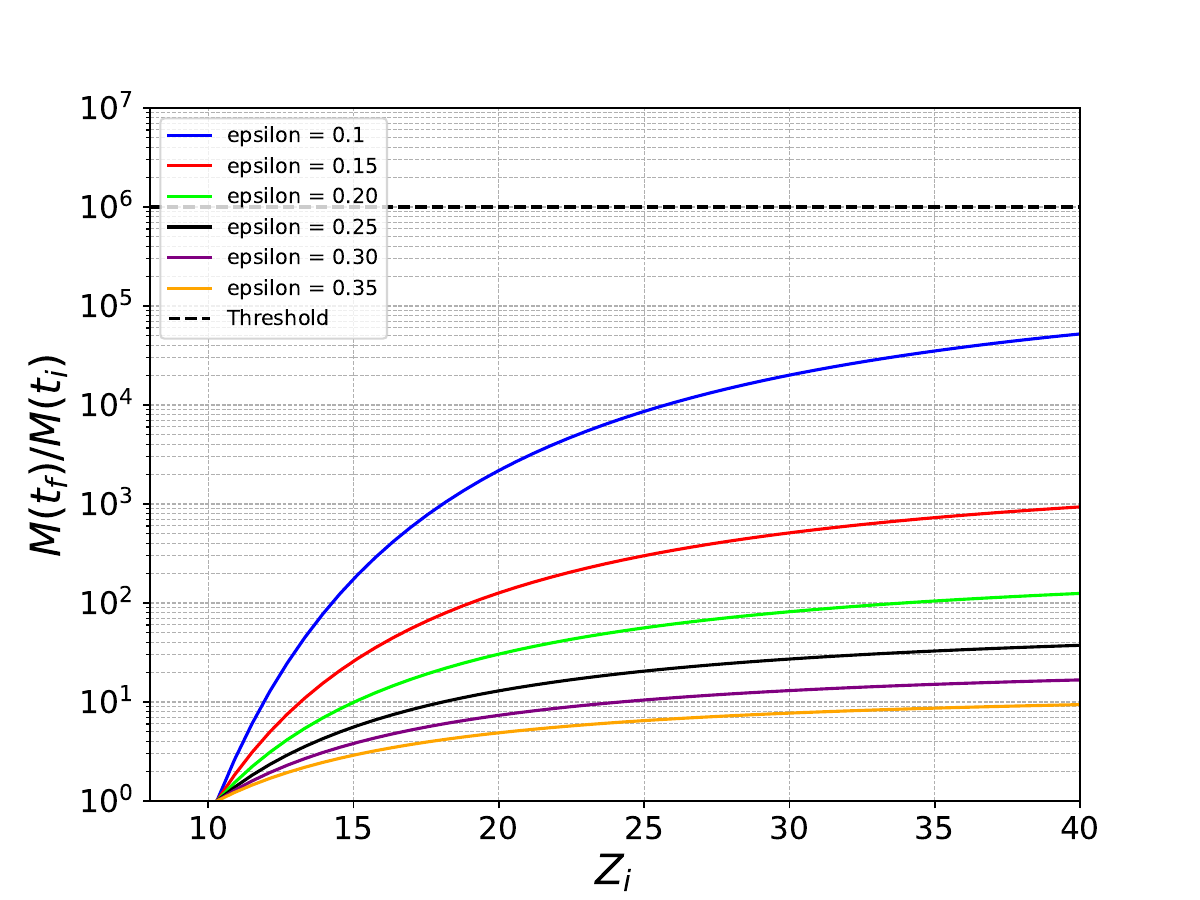}
\caption{Black hole accretion mass amplification $M_f/M_i$ versus redshift $z_i$ of the initial seed. Here we plot the amplification achieved by redshift $z_f$ = 10.3 for $f_{Edd} = 1.0$, and different values of $\epsilon$. The horizontal dashed line indicates the observed mass threshold for a $100M_{\odot}$ seed.}
\label{mass_ampl}
\end{figure}

\begin{figure}
\centering
\includegraphics[scale=0.75]{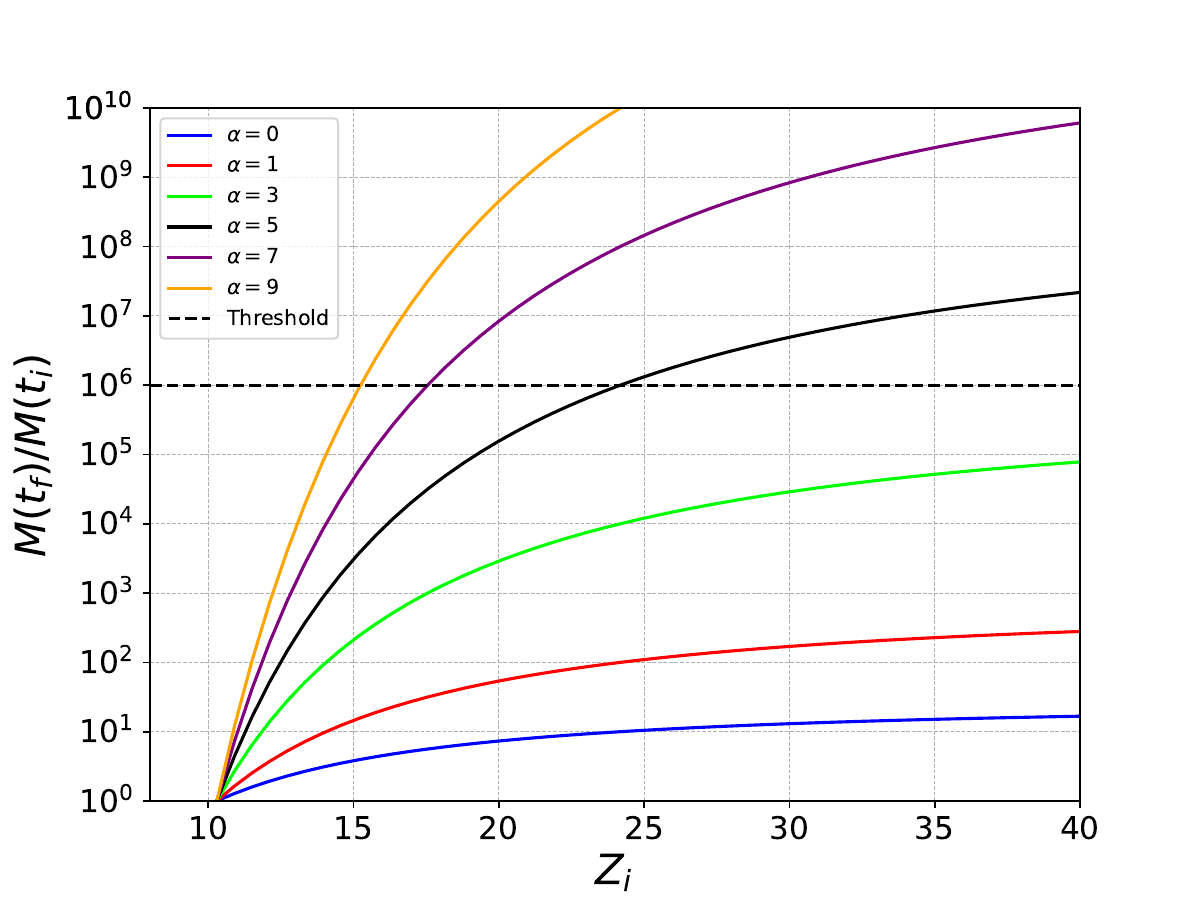}
\caption{Black hole accretion mass amplification $M_f/M_i$ versus redshift $z_i$ of the initial seed. Here we plot the amplification achieved by redshift $z_f$ = 10.3 for $\epsilon = 0.3$, $f_{Edd} = 1.0$, and different values of $\alpha$. The horizontal dashed line indicates the observed mass threshold for a $100M_{\odot}$ seed.}
\label{mass_ampl-mog}
\end{figure}

To study
the black hole mass amplification factor $M_f/M_i$ as a
function of the redshift of the initial seed black hole, we replace the time parameter of Eqs.(\ref{mog-mass-rate}, \ref{mass-rate}) with Eq.(\ref{z}) and plot them in Figs.(\ref{mass_ampl}, \ref{mass_ampl-mog}) in which we consider $z_f = 10.3$ to be the redshift of the highest known quasar detected by the JWST telescope~\cite{2023arXiv230515458B}. To have a comparable result to~\cite{2005ApJ...620...59S}, we plot the black hole mass amplification factors up to $z_i = 40$. In Fig.(\ref{mass_ampl}), we have plotted the mass amplification factor for $f_{Edd} = 1.0$ and for different values of $\epsilon$. As can be seen, for none of the reported $\epsilon$, a seed of $100M_{\odot}$ mass cannot reach a final mass-threshold of $10^{8}M_{\odot}$ at $z=10.3$. We have provided the mass amplification factor for $\epsilon=0.3$, $f_{Edd} = 1.0$, and different values of $\alpha$ for MOG in Fig. (\ref{mass_ampl-mog}). Clearly, the mass threshold of $10^{8}M_{\odot}$ (which is equivalent to $10^{6}$ mass amplification for a $100M_{\odot}$ seed) can be achieved for various values of $\alpha$ even though we have used $\epsilon=0.3$.

\section{Conclusions}\label{section:conclusions}

We have discussed the mass accumulation problem of SMBHs within 800 Myr after the Big Bang and considered some of the proposed solutions for this problem. Alternatively, we showed that MOG can be one of the possible solutions to explain how SHBHs can accumulate such a huge mass at redshifts $z \ge 7$. Considering that the $\alpha$ parameter of the theory strengthens the gravitational force, we showed how SMBHs can gain a mass of order $10^9$ $M_{\odot}$. In Figs.(\ref{alpha-eps}, \ref{alpha-fedd}) we demonstrated that to have SMBHs with a mass of the order $10^{9}M_{\odot}$, there is a degeneracy in choosing $\alpha$, $\epsilon$ and $\fedd$, which need to be constrained by future observations.
Finally, we calculated the mass amplification factor and plotted it as a function of redshift in Figs.(\ref{mass_ampl}, \ref{mass_ampl-mog}) for the concordance model and MOG SMBH. We demonstrated that although standard SMBHs are unable to have big enough mass amplification factors for reasonable values of $\epsilon$,  MOG SMBHs can provide such mass amplification factors that can explain observations.

\section{ACKNOWLEDGMENTS}
We thank Niayesh Afshordi for his useful comments, discussions, and support. We also thank Martin Green and Viktor Toth for helpful discussions.
This research was supported in part by Perimeter Institute for Theoretical Physics. Research at Perimeter Institute is supported by the
Government of Canada through the Department of Innovation, Science, and Economic Development Canada and by the Province of Ontario through
the Ministry of Research, Innovation and Science.

\bibliography{biblio} 

\end{document}